\documentclass[prl,twocolumn,showpacs,preprintnumbers,amsmath,amssymb,floatfix,superscriptaddress]{revtex4}

\usepackage{graphicx}
\usepackage{dcolumn}
\usepackage{bm}
\usepackage{color}
\usepackage{url}
\usepackage{amsmath}
\usepackage{bm,multirow}

\begin{document}

\newcommand{\Cardiff}{School of Physics and Astronomy, Cardiff University, Queens Building, CF24 3AA, Cardiff, United Kingdom}
\newcommand{\UIB}{Departament de F\'isica, Universitat de les Illes Balears, 
Crta. Valldemossa km 7.5, E-07122 Palma, Spain}

\title{
A simple model of complete precessing black-hole-binary gravitational waveforms
}

\author{Mark Hannam}
\affiliation{\Cardiff}
\author{Patricia Schmidt}
\affiliation{\Cardiff}
\author{Alejandro Boh\'e}
\affiliation{\UIB}
\author{Le\"ila Haegel}
\affiliation{\UIB}
\author{Sascha Husa}
\affiliation{\UIB}
\author{Frank Ohme}
\affiliation{\Cardiff}
\author{Geraint Pratten}
\affiliation{\Cardiff}
\author{Michael P\"urrer}
\affiliation{\Cardiff}

\begin{abstract}
The construction of a model of the gravitational-wave (GW) signal from generic configurations of spinning-black-hole
binaries, through inspiral, merger and ringdown, is one of the most pressing theoretical problems in the build-up 
to the era of GW astronomy. We present the first such model in the frequency domain, ``PhenomP'', which captures the basic phenomenology of the 
seven-dimensional parameter space of
binary configurations with \emph{only three} key physical parameters. Two of these (the binary's mass ratio and
an effective total spin parallel to the orbital angular momentum, which determines the inspiral rate) define an 
underlying non-precessing-binary model. The non-precessing-binary waveforms are then ``twisted up'' 
%Essentially, we ``twist up'' a two-parameter 
%non-precessing-binary model 
with approximate expressions for the precessional motion,
which require only one additional physical parameter, an \emph{effective precession spin}, $\chi_p$. 
All other parameters (total mass, sky location, orientation and polarisation, and initial phase) can be specified trivially.
The model is constructed in the frequency domain, 
which will be essential for efficient GW searches and source measurements. We have tested 
the model's fidelity for GW applications by
comparison against hybrid post-Newtonian--numerical-relativity waveforms at a variety of 
configurations --- although we \emph{did not use these numerical simulations} 
 in the construction of the model. 
 Our model
can be used to develop GW searches, to study the implications for astrophysical
measurements, and as a simple 
conceptual framework to form the basis of generic-binary waveform modelling in the advanced-detector
era. 
\end{abstract}
\maketitle

\paragraph{Introduction.---}

The imminent commissioning of second-generation laser-interferometric gravitational-wave
detectors will bring us closer to the era of gravitational-wave (GW) astronomy, which carries
the potential to revolutionize our understanding of astrophysics, fundamental physics, and 
cosmology~\cite{lrr-2009-2}. Among the most promising GW sources are the inspiral and merger 
of black-hole binaries. Detection and interpretation of these signals requires analytic models that
capture the phenomenology of all likely binary configurations; most of these will include 
complex precession effects due to the black-hole spins. However, most of the current models
of the two black holes' inspiral, their merger, and the ringdown of the final black hole, consider
only configurations where the black-hole spins are aligned with the binary's orbital angular 
momentum, which \emph{do not} involve any precession. 

The binary's early inspiral can be modelled with analytic 
post-Newtonian (PN) calculations, but the late inspiral and merger require 3D numerical solutions
of the full nonlinear Einstein equations. These expensive numerical relativity (NR) calculations
must span a parameter space of binary configurations that covers, for non-eccentric inspiral,
seven dimensions: the mass ratio of the binary, and the components of each black hole's 
spin vector;
the total mass of the system is an overall scale factor. 
A naive mapping with at least four simulations in each direction of parameter space (as was 
sufficient for the current phenomenological non-precessing 
models~\cite{Ajith:2007kx,Ajith:2009bn,Santamaria:2010yb}) would imply that modelling these
systems requires $4^7 \sim O(10^4)$ numerical simuations.
%Previous work on 
%phenomenological
%models of non-precessing binaries suggests that we require at least four simulations in each
%direction of parameter space that we intend to 
%model~\cite{Ajith:2007kx,Ajith:2009bn,Santamaria:2010yb}. This implies that we need 
%$4^7 \approx 16,000$ 
%numerical simulations to model the full parameter space, which is unfeasible in the near future. 
%We must identify approximations and degeneracies that make the task tractable. 
%\mdh{The referee objects to this statement, and of course it is only an estimate based on the 
%phenom work. However, the current non-precessing EOB model is calibrated to the same number
%of waveforms as the Phenom B and C models (27), so the claim is not limited to Phenom. Going into
%this detail seems too much. Maybe it is better to put it a bit softer, like:
%``A naive mapping with at least four simulations in each direction of parameter space (as was 
%sufficient for the current phenomenological non-precessing 
%models~\cite{Ajith:2007kx,Ajith:2009bn,Santamaria:2010yb}) would imply $4^7 \approx 16,000$ 
%numerical simuations.'' ??}

In recent work we identified an approximate mapping between inspiral waveforms from generic 
binaries, and those from a \emph{two-dimensional} parameter space of non-precessing 
binaries~\cite{Schmidt:2012rh}. 
This approximation holds because precession has little effect on the inspiral rate, and so 
precession effects approximately decouple from the overall inspiral, which can be described by
a non-precessing-binary model, neglecting the effect of breaking equatorial symmetry, which is 
responsible for large recoils~\cite{Brugmann:2007zj}. 
We further proposed that, given a model for the precessional motion 
of a binary, we could construct an approximate waveform by ``twisting up'' the appropriate 
non-precessing-binary waveform with the precessional motion. This technique was recently 
adopted to produce simple frequency-domain PN inspiral waveforms~\cite{Lundgren:2013jla}. 
It was more recently suggested that this mapping also holds through merger and 
ringdown~\cite{Pekowsky:2013ska}.  

In this work we take 
this idea further, in two crucial ways. 
First, we use PN expressions for the precession 
angles to twist up a phenomenological 
model of non-precessing-binary waveforms~\cite{Santamaria:2010yb}, which includes merger and ringdown.
The inclusion of merger and ringdown provides the first 
frequency-domain inspiral-merger-ringdown model of 
generic binaries. (Frequency-domain models are essential for both efficient GW searches and 
parameter estimation.)
Our model uses the highest-order (closed-form) PN expressions available, and we 
also incorporate precession effects into the estimate of the final black-hole spin and
the ringdown model. 

Second, we make use of a \emph{single} parameter that captures the basic precession phenomenology
for \emph{generic} binary configurations~\cite{Schmidt:2014iyl}. Our final model has only \emph{three} 
intrinsic dimensionless physical parameters, the two parameters of our previous non-precessing models 
(the mass ratio $q=m_2/m_1 \geq 1$, and an effective \emph{inspiral} spin, $\chi_{\rm eff}$, 
which characterizes the rate of inspiral); plus one additional parameter, an \emph{effective precession spin}, 
$\chi_p$. All other additional configuration parameters (the total mass, the binary's initial orbital phase and
initial precession angle, plus its sky location, orientation, polarisation, and initial orbital and precession phases), 
can be trivially included analytically. We describe this parameterization in more detail below; 
its effectiveness in capturing the 
phenomenology of the inspiral across the full parameter space is
demonstrated in Ref.~\cite{Schmidt:2014iyl}. Our evaluation of its fidelity for GW applications 
when including merger and ringdown by comparison against hybrid PN-NR waveforms  constitutes
our core quantitative result.

The purpose of this model is to (a) facilitate the development of computationally efficient 
generic-binary searches, (b) provide a starting point to investigate the parameter-estimation 
possibilities (and limitations) of generic-binary observations in second-generation detectors, 
and their astrophysical implications, and (c) as a simple framework for the construction of more refined models 
calibrated to NR simulations. 
If the dominant parameter space of binary simulations can be reduced to three dimensions
(mass ratio, effective inspiral spin, effective precession spin), it may be feasible to produce 
a sufficient number of NR waveforms ($\sim$100) to calibrate the model well before advanced 
detectors reach design 
sensitivity in 2018-20~\cite{Aasi:2013wya}. 
The model can be further refined, based on the results of these studies. 
As such, this model provides a practical road map to model generic binaries 
to meet the needs of GW astronomy over the next decade. 
This model has been included in the LAL data analysis software, to facilitate the
development and testing of search and parameter estimation pipelines~\cite{LAL}.

\paragraph{Model.---}

% \begin{figure}
%\begin{center}
%  \includegraphics[width=0.4\textwidth]{Coordinates}
%  \caption{
%  \label{fig:coordinates} 
%The precession angles $(\iota,\alpha)$ of the Newtonian orbital angular momentum
%$\hat{\bm L}$ about the total 
%angular momentum $\hat{\bm J}$. The figure also indicates one early precession cycle
%for a $q=3$ hybrid PN-NR
%waveform (red), and the results from the PN formulas used in the PhenomP model (black). The different
%curve lengths indicate the dephasing in $\alpha$; see text.
%    }
%\end{center}
%\end{figure}
% 
% MDH: Removed figure. Can instead refer to a figure in some other paper. 

We start from the frequency-domain model (``PhenomC'')~\cite{Santamaria:2010yb} 
of non-precessing waveforms, because it includes the 
standard state-of-the-art inspiral phase. This model describes the $(\ell= 2, m =|2|)$ modes of the 
waveform, with $h(f)  =  A(f) e^{i \psi(f)}$, where $A(f)$ and $\psi(f)$ are given in Ref.~\cite{Santamaria:2010yb}. 
Based on the approximate mapping identified in Ref.~\cite{Schmidt:2012rh}, for a given generic binary 
we start with the non-precessing waveform given by the parameters $(M, \eta, 
\chi_{\rm eff})$, where $\eta = q/(1+q)^2$ and $\chi_{\rm eff} = (m_1 \chi_1 + m_2 \chi_2)/M$;
$\chi_1$ and $\chi_2$ are the components of the dimensionless spins ($\chi_i = \bm
S_i \cdot \hat{\bm L}/m_i^2$) projected along the Newtonian orbital angular momentum $\hat{\bm L}$. 
The direction of $\hat{\bm J}$ is approximately constant throughout the evolution, as angular-momentum loss 
via GWs is predominantly along $\hat{\bm J}$, with emission orthogonal to $\hat{\bm J}$ averaging out due to the 
precession of $\hat{\bm L}$ around $\hat{\bm J}$ ~\cite{Apostolatos:1994mx}. We therefore assume that the final spin is 
in the same direction as $\hat{\bm J}$ through the inspiral, and update the PhenomC final spin 
magnitude estimate~\cite{Berti:2005ys} to account for precession, using Ref.~\cite{Barausse:2009uz}, 
with only one black hole spinning.

We then twist up the non-precessing model, i.e., we approximate the $\ell=2$ modes of a precessing 
binary waveform in the time domain by rotating the dominant modes of the corresponding non-precessing 
waveform~\cite{Schmidt:2010it,Schmidt:2012rh} as
\begin{equation}
 h_{2m}^P(t) = e^{-i m \alpha} \sum_{|m'|=2} e^{ i m' \epsilon} d^2_{m',m}(-\iota) h_{2,m'}(t),
\label{eqn:trans}
\end{equation}
where $d^\ell_{nm}$ denote the Wigner d-matrices. The angles $\alpha$ and $\iota$ that enter our model are 
defined as the spherical angles parametrizing 
the unit Newtonian orbital angular momentum $\hat{\bm L}$ 
(see, for example, Fig.~1 in Ref.~\cite{Schmidt:2014iyl})
in an inertial frame with 
$\hat{\bm z}=\hat{\bm J}$. The third angle, defined from
$\dot{\epsilon}=\dot{\alpha} \cos \iota$, parametrizes 
a rotation around $\hat{\bm L}$ \cite{Boyle2011}. During the inspiral phase, all of these angles vary slowly 
(on the precession timescale) with respect to the orbital timescale, which allows for a stationary-phase-approximation 
(SPA) transformation to the frequency domain (this fact has been exploited in work dating from 
Ref.~\cite{Apostolatos:1996}, and was most recently used in Ref.~\cite{Lundgren:2013jla}). 
Here, we use closed-form frequency-domain PN expressions for these angles (valid for systems with only one spin in the 
orbital plane) to twist the entire non-precessing modes, formally continuing the SPA treatment through merger and ringdown. 
Although we do not expect these expressions, or the approximation of slowly varying precession angles, 
to be valid through merger and ringdown, in practice we find that
they mimic to reasonable accuracy the phenomenology of our PN-NR hybrids and lead to high  
fitting factors even for high masses. Our model consists entirely of closed-form analytic expressions, and
the output is the two polarizations $h_{+,\times}^P(Mf;\eta,\chi_{\rm eff},\chi_p,\theta,\phi)$.

The inclination $\iota$ is simply the angle between the binary's total angular
momentum, $\hat{\bm J}$, and orbital angular momentum,
$\hat{\bm L}$, so that  $\cos \iota  = \hat{\bm L} \cdot  \hat{\bm J}=\hat{\bm L} \cdot \bm J/|\bm J|$. 
In practice we find that the accuracy in $\iota$, which enters only in amplitude factors for the contributions
 in \eqref{eqn:trans}, is not critical and that it is sufficient to include only non spinning corrections in $\bm J$ 
 beyond the total spin contribution at leading order.  The precession angle $\alpha$ is computed using the 
 expression for $\dot{\alpha}$ obtained in \cite{Blanchet2011} (see Eqs (4.10a) and (4.8)) by plugging in the 
 highest order (next-to-next-to-leading in spin-orbit) expressions available for the quantities entering the 
 formula \cite{Bohe2013}, PN re-expanding and averaging over the orientation of the spin in the orbital plane.

The spin parameters in our model are $\chi_{\rm eff}$ and $\chi_p$. The effective inspiral spin
$\chi_{\rm eff}$  was
defined earlier. The angle expressions $(\alpha, \iota)$, require some choice for the 
distribution of 
spins across the two black holes, and for our implementation we let $\chi_1 = 0$ and $\chi_2 = (M/m_2) \chi_{\rm eff}$,
i.e., all of the spin is on the larger black hole. This choice performs well in the study in Ref.~\cite{Schmidt:2014iyl}.
To ensure \emph{physical} spins of $\chi \leq 1$ for
each black hole, we could also choose $\chi_1 = \chi_2 = \chi_{\rm eff}$. The implications of these
choices for detection and parameter estimation will be explored in future work; in the cases
we study here, we see that our model is likely to perform well for GW detection.
The in-plane spin magnitude $\chi_p$ is associated with the larger black hole. 

We expect our model to capture the basic phenomenology of generic two-spin systems, motivated
by the following argument. 
For the effective precession spin,
if $S_{1\bot}$ and $S_{2\bot}$ are the magnitudes of the 
projections of the two spins in the orbital plane, then, 
according to the PN precession equations~\cite{Apostolatos:1994mx, Kidder:1995zr}, 
the precession rate at leading order will be proportional to 
$(A_1 S_{1\bot} + A_2 S_{2\bot})$ when the vectors $\bm S_{1\bot}$ and
$\bm{S}_{2\bot}$ are parallel, and
by $(A_1 S_{1\bot} - A_2 S_{2\bot})$ when they point in opposite directions,
where $A_i = 2 + (3 m_{3-i})/(2 m_{i})$. 
During the inspiral, to first approximation the average precession rate for non-equal-mass systems 
is simply the maximum of these two spin contributions, and we
can define $S_p = \max(A_1 S_{1\bot}, A_2 S_{2\bot})/A_2$,
and expect that applying an
in-plane spin of $\chi_p = S_p / m_2^2$ to the larger black hole will mimic the main precession effects of the 
full two-spin system. In equal-mass systems a double-spin configuration is indistinguishable from a 
single-spin system; this changes the interpretation of $\chi_p$, but the use of two spin parameters is now
automatically valid. (The third spin-vector component corresponds to the initial precession angle of the system,
which is included as an overall complex factor in the model.) Full generic two-spin waveforms 
 will typically exhibit additional small oscillations in the precession angles 
(see e.g.,~Fig.~4 of Ref.~\cite{Buonanno:2004yd}, and
Ref.~\cite{Schmidt:2014iyl}), but we do not expect these effects to be detectable in most GW observations.
These two parameters, $\chi_{\rm eff}$ and $\chi_p$, can be mapped to a range of
physically allowable individual black-hole spins.

\paragraph{Results.---}

The most reliable way to test our model is to compare against hybrid PN (inspiral) and NR (merger-ringdown)
waveforms. But to do that across the full generic-binary parameter space would 
require the same number of waveforms as needed to construct a seven-dimensional generic model, 
which is the computationally prohibitive task that we wished to avoid in the first place. In practice all we can
do is identify what we expect to be challenging points in the parameter space. In this work we restrict ourselves
to binaries with mass ratios $q \leq 3$, because that is the mass ratio to which the underlying PhenomC model
was calibrated to spinning-binary waveforms. 
We construct four hybrids at mass ratios 2 and 3, for
a variety of spin choices. The numerical simulations were produced with the BAM 
code~\cite{Brugmann:2008zz}, and hybrids
were constructed by the method described in Ref.~\cite{Schmidt:2012rh}, and also in the inertial frame of the 
NR waveforms, for comparison. 
The comparison configurations are chosen to include strong precession (a $q=3$ case where the 
larger black hole has a spin of $\chi_2 = 0.75$ in the orbital plane), a double-spin $q=2$ case, where the
small BH has spin 0.5, and the larger black hole spin 0.75, both in the orbital plane, which tests our 
assumption that we can consider only a weighted average of the spins when constructing $\chi_p$. 
We also consider a double-spin $q=3$ case where $\chi_{\rm eff} = -0.5$
and $\chi_p = 0.6$, and another where $\chi_{\rm eff} = -0.125$ and $\chi_p = 0.75$; the purpose here was
to test the model with non-zero $\chi_{\rm eff}$. The NR waveforms include between 10 and 14 GW cycles
before merger, and the PN part consists of $\sim$200 cycles. For all waveforms we considered, there are
approximately three pre-merger precession cycles, since the precession rate depends only weakly on the
mass ratio.

\begin{figure}
\begin{center}
  \includegraphics[width=0.45\textwidth]{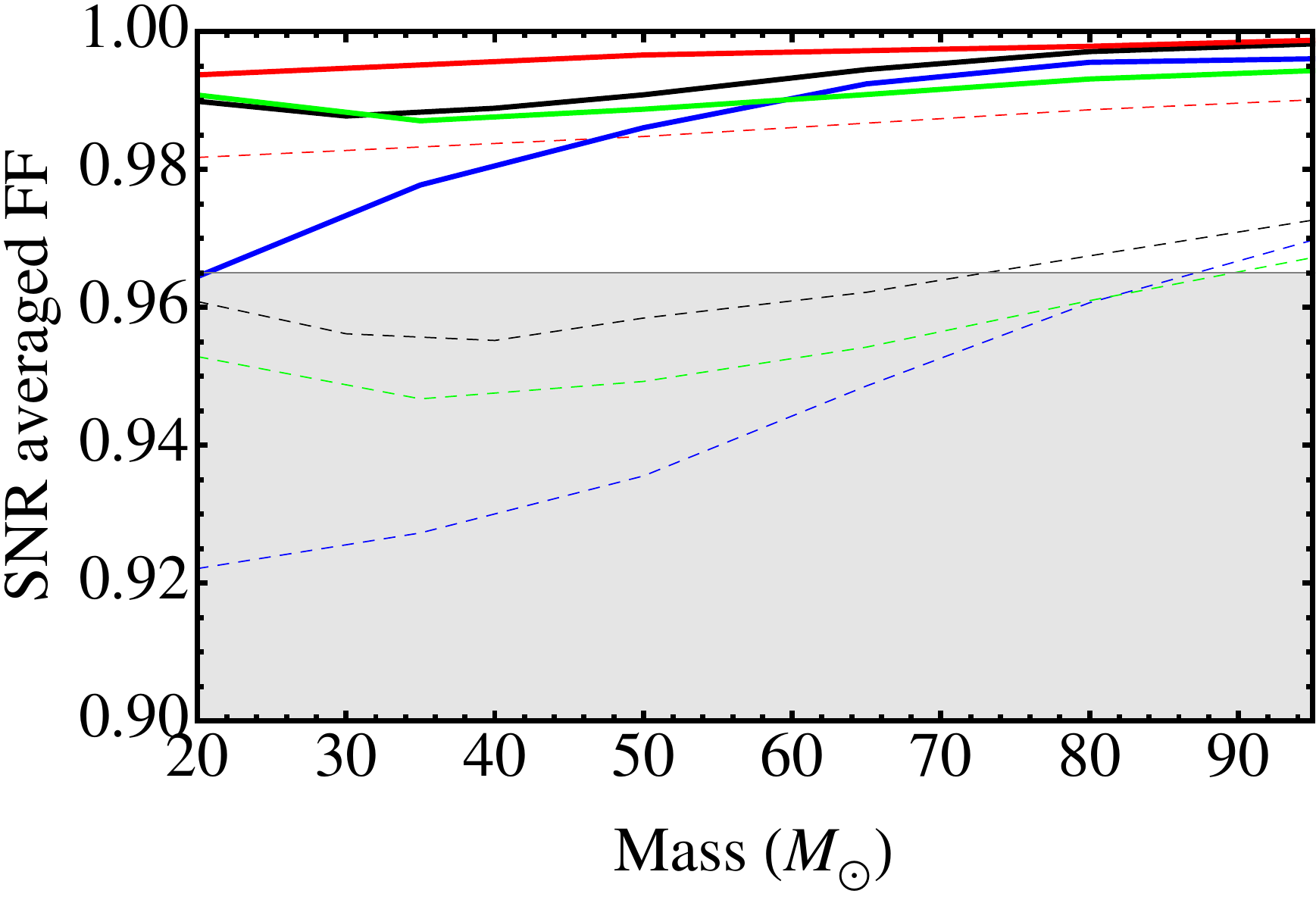}
  \caption{
  \label{fig:AverageFF} 
  Fitting factors between PhenomP (solid lines) and PhenomC (dashed lines), averaged over binary orientations, 
  as described in the text. Each color is one case:
  $q=3$, $\chi_p = 0.75$ (black); $q=2$ double spin (red); 
 $q=3$, $\chi_{\rm eff} = -0.5$ (green);
 $q=3$, $\chi_{\rm eff} = -0.125$ (blue). We see that in all cases PhenomP meets the 0.965 threshold
 for detection accuracy. Above 100\,$M_\odot$ all of the curves are above 0.965.
    }
\end{center}
\end{figure}

\begin{figure}
\begin{center}
  \includegraphics[width=0.35\textwidth]{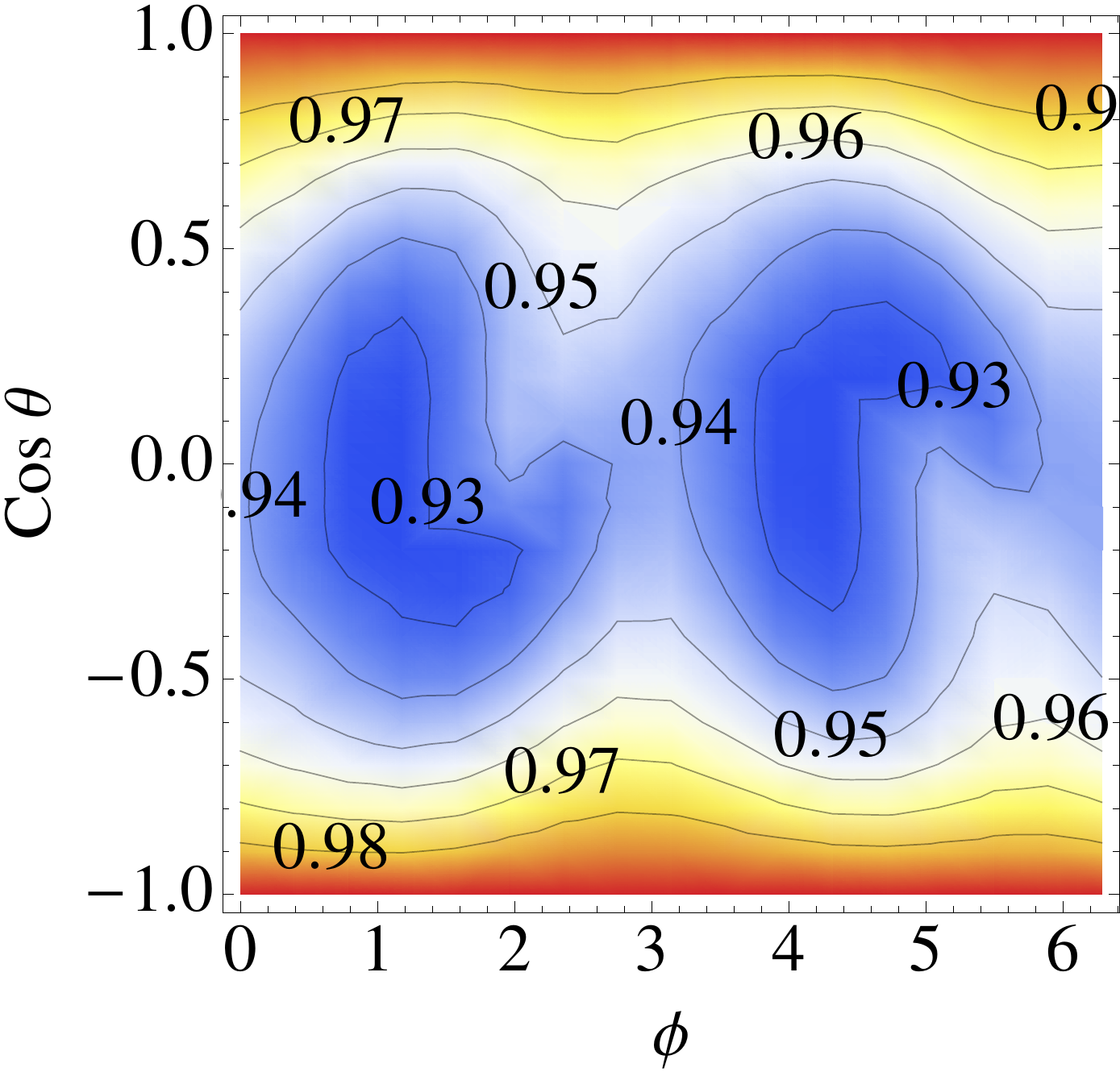}
  \includegraphics[width=0.35\textwidth]{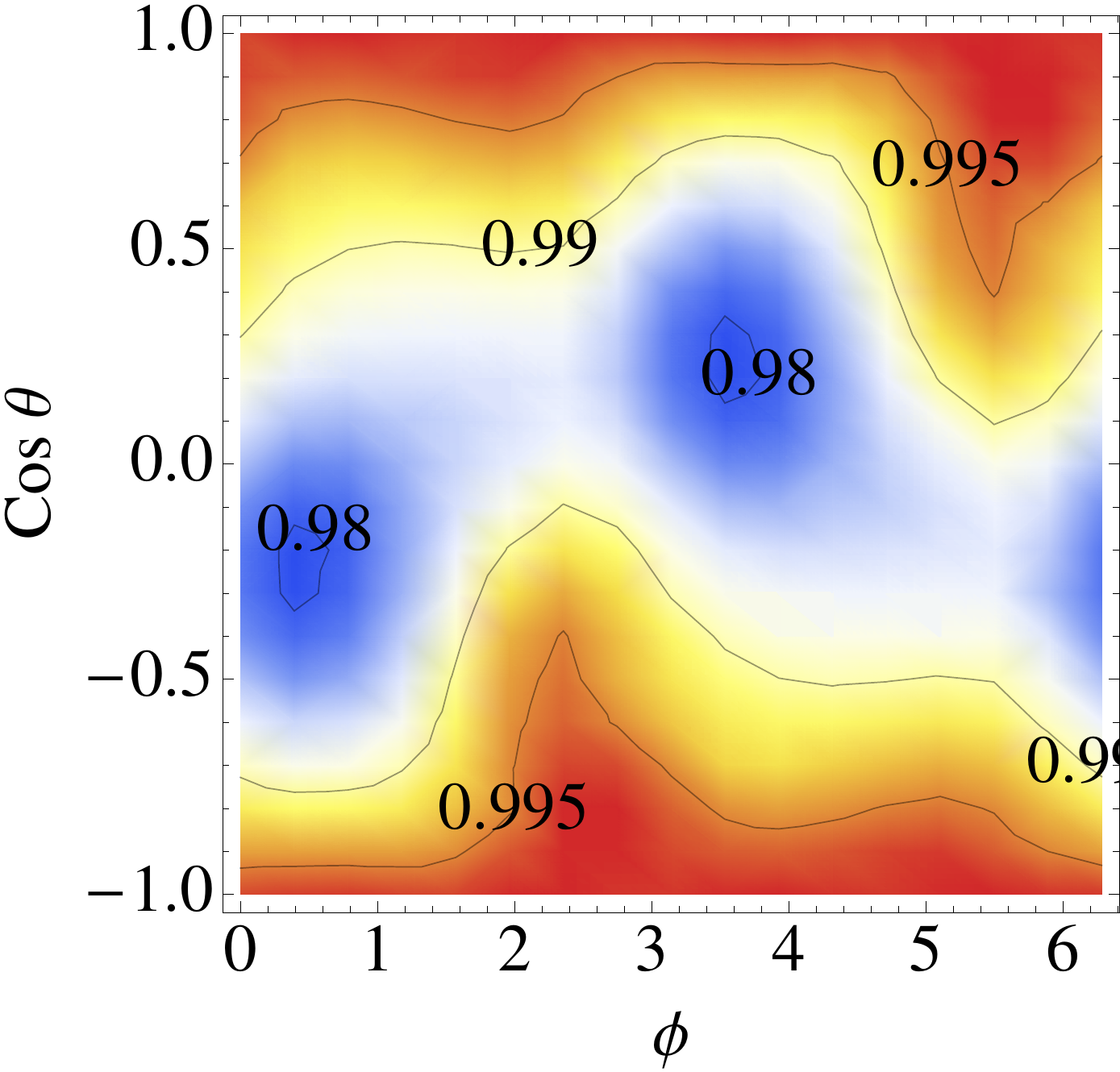}
  \caption{
  \label{fig:FF} 
  Fitting factors (FF) between a $q=3$ highly-precessing binary, and the non-precessing PhenomC and
  precessing PhenomP models, as a function of binary orientation angles $(\theta, \phi)$; at $\theta = 0$
  an observer is oriented with the binary's
   total angular momentum. FF $<0.965$ for many orientations with PhenomC, while for PhenomP 
  it is well above 0.965 for all orientations. See text for further details.
    }
\end{center}
\end{figure}

As is standard in GW analysis, we
calculate the noise-weighted inner product between our source waveform (in this case the hybrid), and a model
(either the original non-precessing PhenomC model, or our new precessing
``PhenomP'' model). We use the current expectation for the design sensitivity of advanced LIGO~\cite{T0900288}, 
with a low-frequency cutoff of 20Hz. This inner product
is maximised with respect to the parameters of the model, including the physical parameters and the binary orientation
and polarization. This optimised inner product is called the ``fitting factor''; its value indicates how well the signal 
can be found in detector data, and the bias between the best-fit model parameters, and the true source 
parameters, give us an indication of the errors in a GW measurement.
We have computed fitting factors using PhenomC and PhenomP for total source masses between 
20\,$M_\odot$ and 200\,$M_\odot$ and as functions of binary orientations.

Fig.~\ref{fig:AverageFF} shows the fitting factor averaged across binary orientations, appropriately weighted with the 
signal-to-noise ratio to give an indication of the proportion of signals that would be detected. 
The standard requirement for GW searches is that the fitting factor be above 0.965, 
corresponding to a loss of no more than 10\% of sources in a search
(disregarding additional loss due to a discrete template bank). We see that in all cases the PhenomP model 
achieves average fitting factors above this threshold. The PhenomC model is acceptable only at high masses. 

The fitting factors are highest for near-optimal orientations, where the total angular momentum is aligned with 
the detector, and from which the precession has only a small effect
on the signal. As an example, Fig.~\ref{fig:FF} shows results for the $q=3$ high precession configuration 
at 50\,$M_\odot$, which proved to be the most challenging configuration with no parallel spin component. 
Results are similar for lower masses, while for higher masses fitting 
factors improve at the expense of parameter accuracy. For the case with large negative $\chi_{\rm eff}$ we
found that for some orientations at masses before $60\,M_\odot$, the fitting factor was below the 0.965 
threshold. This is because of differences in 
%Comparing the two panels of Fig.~\ref{fig:FF}  we see that while the fitting factors for
%PhenomC are above 0.97 only for near-optimal orientations (from which the precession has only a small effect
%on the signal), they are above 0.97 for almost all orientations with the PhenomP model. 
%For the cases with $\chi_{eff} \neq 0$ we also saw good fitting factors at high masses $M > 50\,M_\odot$, 
%while at low masses the fitting factors deteriorated due to differences in
 the PN inspiral approximants used
for the hybrid and for the model; these effects for large anti-aligned spins have been observed in the
past~\cite{Nitz:2013mxa}, and are independent of our modelling procedure. 
We leave a full 
study of parameter biases to future work, but our results suggest that a measurement of $\chi_p$ reliably 
identifies precession.

\paragraph{Discussion.---}

We have presented the first frequency-domain inspiral-merger-ringdown model for the GW signal from 
precessing-black-hole binaries. Incorporating a series of insights from our previous work,
our model is constructed by a straightforward transformation of a non-precessing-binary model, in this
case PhenomC; in practice any workable non-precessing model
could be used instead. The current model did not require any precessing-binary numerical simulations in its 
construction, although in the future we plan to use extensive simulations to refine the model, based on 
tests of the model's accuracy for GW searches and parameter estimation. Finally, we are able to 
model the essential phenomenology of the seven-dimensional parameter space of binary 
configurations with a model that requires only three key \emph{physical} parameters. This will simplify the model's 
incorporation into search and parameter estimation pipelines, as well as making tractable the problem
of producing enough numerical simulations to produce a model of sufficient accuracy for GW
astronomy with advanced detectors.
It is not clear whether a precessing IMR model will be necessary for GW searches, or the level of accuracy
that is required for parameter estimation studies, but these questions can only be answered once a 
(resasonably fast to evaluate) model exists, and we have provided one. 

Our ability to model generic waveforms with only two spin parameters implies strong
degeneracies that will make it difficult to identify the individual black-hole spins, in particular the spin of
the smaller black hole. This may well 
be the reality of GW observations with second-generation detectors, for which 80\% of signals will 
be at signal-to-noise ratios between 10 and 20, in which the subtle double-spin effects on the 
waveform may be difficult to identify. 
These are important issues that deserve further attention in
future work.

The current model is valid only in the region of parameter space for which PhenomC
was calibrated ($q \leq 3$, $|\chi_{\rm eff}| \leq 0.75$). More challenging precession cases are 
expected at higher mass ratios and spins (e.g., transitional precession), and the ability of our
prescription to model those configurations will need to be tested when refined frquency-domain
non-precessing-binary models become available. 

\paragraph{Acknowledgements.---}

We thank P. Ajith and S. Fairhurst for useful discussions. 
PS is a recipient of a DOC-fFORTE-fellowship of the Austrian Academy of Sciences
and was also partially supported by the STFC.
MH was supported by STFC grants ST/H008438/1 and ST/I001085/1, and FO
and MP  
by ST/I001085/1.
AB and SH were supported by
the Spanish MIMECO grants FPA2010-16495 and CSD2009-00064, European Union FEDER funds, and  
Conselleria d'Economia i Competitivitat del Govern de les Illes Balears.
GP was supported by an STFC doctoral training grant.
LH was supported by the Conseil G\'en\'eral de l'Essonne.
Numerical simulations were carried out at  ARCCA in Cardiff, the UK DiRAC clusters,
MareNostrum at Barcelona Supercomputing Center -- Centro Nacional de 
Supercomputaci\'on, and on the 
PRACE clusters Hermit, Curie and SuperMUC.

\bibliography{phenomp}

\end{document}